\documentclass{article}
\usepackage{amssymb}
\usepackage{amsmath}
\usepackage{latexsym}
\usepackage{array}
\usepackage{graphicx}
\usepackage{bm}
\usepackage{exscale}

\usepackage{theorem}
\theoremstyle{break}
\newtheorem{Theorem}{Theorem}
\newtheorem{Proposition}[Theorem]{Proposition}
\newtheorem{Lemma}[Theorem]{Lemma}
\newtheorem{Definition}{Definition}
\def\qed{\hfill\hbox{$\Box$}\vspace{10pt}\break}

\makeatletter
\def\maketitle{%
\begin{flushright}%
{\small \@date}
\end{flushright}%
\begin{flushleft}%
{\Large \@title \par}
\end{flushleft}%
\begin{flushleft}%
{\normalsize \@author}
\end{flushleft}%
\par\vskip 1.5em
\@thanks
}
\makeatother

\begin{document}
\title{\bf Irreducibility and co-primeness as an integrability criterion for discrete equations}
\author{\bf Masataka Kanki$^1$, Jun Mada$^2$, Takafumi Mase$^3$ and\\
Tetsuji Tokihiro$^3$\\
\rm \small $^1$ Department of Mathematics, Faculty of Science\\
\small Rikkyo University, 3-34-1 Nishi-Ikebukuro, Tokyo 171-8501, Japan\\
\small $^2$ College of Industrial Technology,\\
\small Nihon University, 2-11-1 Shin-ei, Narashino, Chiba 275-8576, Japan\\
\small $^3$ Graduate School of Mathematical Sciences,\\
\small University of Tokyo, 3-8-1 Komaba, Tokyo 153-8914, Japan}
\date{}
\maketitle

\begin{abstract}
We study the Laurent property, the irreducibility and co-primeness of discrete integrable and non-integrable equations.
First we study a discrete integrable equation related to the Somos-4 sequence, and also a non-integrable equation as a comparison. We prove that the conditions of irreducibility and co-primeness hold only in the integrable case.
Next, we generalize our previous results on the singularities of the discrete Korteweg-de Vries (dKdV) equation.
In our previous paper \cite{dKdVSC} we described the singularity confinement test (one of the integrability criteria) using the Laurent property, and the irreducibility, and co-primeness of the terms in the bilinear dKdV equation, in which we only considered simplified boundary conditions. This restriction was needed to obtain simple (monomial) relations between the bilinear form and the nonlinear form of the dKdV equation.
In this paper, we prove the co-primeness of the terms in the nonlinear dKdV equation for general initial conditions and boundary conditions, by using the localization of Laurent rings and the interchange of the axes.
We assert that co-primeness of the terms can be used as a new integrability criterion, which is a mathematical re-interpretation of the confinement of singularities in the case of discrete equations.

{\tt MSC2010: 37K10, 35Q53, 35A20}

{\tt PACS: 02.10.Hh, 02.30.Ik, 02.40.Xx, 45.05.+x}

{\tt Keywords: singularity confinement, discrete KdV equation, Somos sequence, co-primeness}
\end{abstract}

\section{Introduction}
\label{sec1}
This article studies the integrability of several difference equations and its relation to the concepts of the Laurent property, irreducibility and co-primeness.
The idea of using these three concepts for integrability detection of discrete equations has first been introduced in our previous paper \cite{dKdVSC}, and this paper is a follow up from the results there. In this paper, we consider a type of QRT mappings and the discrete KdV equation with a general boundary condition (cf. In \cite{dKdVSC} we treated the equation with a simpler boundary condition).

Let us first briefly review the theory of integrable equations.
We have long been searching for a criterion for integrability of differential and difference equations.
In the case of ordinary differential equations (continuous case), the Painlev\'{e} property is a useful criterion for integrability \cite{Conte}. Motivated by this property,
Grammaticos, Ramani and Papageorgiou introduced the
`singularity confinement' test \cite{SC}, which is now widely used as one of several reliable integrability detectors for ordinary difference equations.
They have discovered that there is a difference between the singularity structures of integrable and non-integrable discrete systems.
In the case of integrable systems, because of a fine cancellation of terms,
even if we start from infinity, we are lead to finite values after a finite number of steps. On the other hand, in non-integrable equations, singularities radiate out from a singular point.
Singularity confinement has been successfully utilized to identify discrete versions of the Painlev\'{e} equations \cite{RGH}.

Motivated by the results of the singularity confinement test, Sakai has completely classified the discrete Painlev\'{e} equations as bi-rational maps on rational surfaces obtained by blowing up two dimensional projective spaces \cite{Sakai}.  
The singularity confinement approach was also taken for partial difference equations in their bilinear forms by Ramani, Grammaticos and Satsuma \cite{RGS}. 
In their approach, singularities are defined as the zeros of the dependent variables and their confinement means that zeros will not propagate in generic cases. 
However, if the space of initial conditions has many independent variables, it is not practical to go through all the configurations of singular points one by one. Thus we need to redefine the confinement of singularities in terms of the algebraic or analytic relations between adjacent terms.

Recently, the integrability of the discrete equations was re-investigated in terms of the `Laurent property': the phenomena that all the terms of the equation are written as Laurent polynomials of the initial variables.
The Laurent property of the Hirota-Miwa equation and the bilinear form of the dKdV equation \eqref{bilin} has been established for generic initial conditions using the theory of cluster algebras \cite{FZ,FZ2}, and this property has been further studied recently for many of the known discrete integrable equations with various initial conditions \cite{Mase}.
In this paper, we first present in section \ref{section2} two examples: one integrable and one non-integrable ordinary difference equation. We prove that the integrable one \eqref{QRT2} has the irreducibility and co-primeness property (theorems \ref{somosirred} and \ref{QRTprop}). The integrable equation we use is a type of QRT mapping \cite{QRT}, and is related to the Somos-4 sequence \cite{Gale}.
On the other hand, the non-integrable one \eqref{nonQRT3} does not have these two properties, despite the fact that its form is similar to that of \eqref{QRT2}.

Next, in section \ref{section3}, we study the dKdV equation.
This section contains several results which are generalizations of the previous paper \cite{dKdVSC}.
The bilinear form of the dKdV equation is given by
\begin{equation}
(1+\delta)a_{m-1}^{n-1} a_m^{n+1} = a_{m-1}^n a_m^n + \delta a_{m-1}^{n+1} a_m^{n-1}, \label{bilin}
\end{equation}
where $m$ and $n$ are integer independent variables.
Initial conditions are given by designating the values for the following set $I_a$, consisting of two rows and one column:
\[
I_a=\{a_m^0,\ a_m^1,\ a_0^n | m\ge 0, n\ge 0\}.
\]
The nonlinear form of the discrete KdV equation is given by
\begin{equation}
\frac{1}{w_{m+1}^{n+1}}-\frac{1}{w_m^n}+\frac{\delta}{1+\delta}(w_m^{n+1}-w_{m+1}^n)=0, \label{nonlin}
\end{equation}
where the initial conditions are given by the set $I_w$:
\[
I_w=\{w_m^0,\ w_1^n | m\ge 1, n\ge 0\}.
\]
These two equations are connected by the relation
\begin{equation}
w_m^n=\frac{a_{m-1}^{n+1} a_m^n}{a_m^{n+1} a_{m-1}^n}. \label{relation}
\end{equation}
In our previous work, we focused not only on the Laurent property but also on the irreducibility of the terms and co-primeness between two terms, and we have proved that the bilinear dKdV equation \eqref{bilin} satisfies the Laurent property, the irreducibility and co-primeness for a specialized initial conditions: $a_m^0=1,a_0^n=1\, (m,n\ge 0)$ \cite{dKdVSC}.
By using this property, we have proved that the nonlinear dKdV equation \eqref{nonlin} with a special boundary condition ($w_0^n=1\, (n\ge 0)$) has a certain `co-primeness' property: i.e., two terms $w_m^n, w_{m'}^{n'}$ of \eqref{nonlin} do not have common factors other than monomials if they are separated by more than one cell ($|m-m'|\ge 2$ or $|n-n'|\ge 2$).
In the previous work we had to choose specialized initial data so that the initial conditions of the two equations \eqref{bilin} and \eqref{nonlin} had monomial relations. This restriction was possible because $a_m^n=1\forall m,\forall n$ is a trivial solution for the bilinear dKdV equation \eqref{bilin}.
Another restriction was that the field of definition of the equations had to be of characteristic zero in the previous paper due to some technical requirements in the proofs.
In this paper, we give a complete proof of the Laurent property, the irreducibility and co-primeness of the bilinear dKdV equation for generic initial conditions over any field.
Using this proposition, we prove our main theorem that the nonlinear dKdV equation has a certain `co-primeness' property (theorem \ref{thm4}) for generic initial conditions.

For general discrete equations, we define the `co-primeness' as the property that ``every pair of two terms of the discrete equation is coprime, except for a finite number of pairs''.
Our assertion is that, for discrete equations, having `co-primeness'  is equivalent to passing the singularity confinement, and, therefore, can be used as an integrability detector.

\section{Co-primeness of QRT mappings} \label{section2}
Let us justify the integrability detection by co-primeness, by means of two simple examples: an integrable equation and a non-integrable one.
The first example is one of QRT mapping (a discrete integrable recurrence relation in one variable) \cite{QRT}, and the other one a non-integrable equation which has a form similar to that of the QRT mapping.
One of the simplest form of the QRT mappings is
\begin{equation}
x_{n+1}=\frac{x_n +1}{x_{n-1}x_n^2}. \label{QRT2}
\end{equation}
We take the initial conditions as $x_0=u,\ x_1=t$, where $t$ and $u$ are independent variables.
The evolution of the equation \eqref{QRT2} is
\begin{eqnarray*}
x_2&=&\frac{1+t}{t^2 u},x_3=\frac{t u (1+t+t^2 u)}{(1+t)^2},x_4=\frac{(1+t)(1+2t+t^2+t u +t^2 u +t^3 u^2)}{u (1+t+t^2 u)^2},\\
x_5&=&\frac{(1+t+t^2 u)\{(1+t)^3 (1+u)+2t^2 u^2+3t^3 u^2 +t^4 u^2+t^4 u^3 \} }{t(1+2t+t^2+t u +t^2 u+t^3 u^2)^2},\cdots.
\end{eqnarray*}
When we look at the factor $(1+t+t^2 u)$, for example, we have 
\[\mbox{ord}_{1+t+t^2 u}(x_3)=1, 
\mbox{ord}_{1+t+t^2 u}(x_4)=-2, \mbox{ord}_{1+t+t^2 u}(x_5)=1.\]
Here ord$_g(f)$ denotes a valuation of $f$ with respect to $g$.
We can also see that $(1+t+t^2 u)$  does not appear as a factor in $x_1,x_2$ and $x_m(m\ge 6)$.
In general, any non-monomial factor has the following property: it appears in the three consecutive terms $x_l,x_{l+1},x_{l+2}$ for some integer $l$, while the other terms $x_m(m\neq l,l+1,l+2)$ do not have this factor.
This fact is presented as theorem \ref{QRTprop} at the end of this section.

Let us point out that the mapping \eqref{QRT2} can be transformed into the so-called Somos-4 sequence:
\begin{equation}
y_{n+2} y_{n-2}=y_{n+1} y_{n-1}+y_n^2,\label{somos4}
\end{equation}
via the following change of variables
\begin{equation}
x_n=\frac{y_{n+3}y_{n+1}}{y_{n+2}^2}. \label{somosrel}
\end{equation}
The Somos-4 sequence is famous for the property that $y_n\in\mathbb{Z}$ for all $n\ge 5$, if we take the initial values $y_1=y_2=y_3=y_4=1$ \cite{Gale}.
More generally, it has been proved that every $y_n$ is a Laurent polynomial of the initial variables \cite{FZ2}, however the irreducibility and co-primeness of the terms have not been studied.
Integrality and the Laurent property for the sequences of this kind are studied, for example, in \cite{HoneSwart}.
Let us state two propositions on the Somos-4 sequence.
We take the initial variables as $y_1=a, y_2=b, y_3=c, y_4=d$, and consider the term $y_n$
in the field of rational functions $\mathbb{C}(a,b,c,d)$.
\begin{Proposition} \label{somosLaurent}
The term $y_n$ of the Somos-4 sequence \eqref{somos4} is a Laurent polynomial of the initial variables $a,b,c,d$:
\[
y_n\in \mathbb{Z}\left[a^{\pm},b^{\pm},c^{\pm},d^{\pm}\right].
\]
\end{Proposition}
For a proof we refer the reader to \cite{FZ2,Mase}.

Before proceeding to the details of the irreducibility and co-primeness, let us fix notations and review some basic algebra to facilitate our proof.
\begin{Definition}
Let $R$ be a commutative ring.
An element $f\in R$ is irreducible if, for every decomposition $f=gh$ $(g,h\in R)$,
at least one of $g$ or $h$ is a unit (reversible element) in $R$.
Two elements $f_1,f_2\in R$ are co-prime if the following condition is satisfied:
If we can write $f_1=g_1 h$, $f_2=g_2 h$ $(g_1,g_2,h\in R)$, then $h$ is a unit in $R$.
\end{Definition}
In the ring $R$ of Laurent polynomials, a unit is a \textit{monomial} of the generators of $R$.
Note that we slightly abuse notation so that the element $x$ of the ring $R$ is called `irreducible', when $x$ is not decomposable, or $x$ itself is a unit of the ring $R$. (A unit is not called irreducible in the usual notation in algebra.)
\begin{Definition} \label{localdef}
Let $R$ be an integral domain (a commutative ring that does not have zero divisors other than $0$). A subset $S$ of $R$ is multiplicative if it satisfies the following:
$1\in S$, $0\not\in S$, $st\in S$ for every $s,t\in S$.
Let $F$ be a quotient field of the ring $R$.
If $S\subset R$ is multiplicative, we can define the localization of $R$ with respect to $S$ as the set
\[
\left\{\frac{a}{s} \Big| s\in S,\, a\in R \right\}.
\]
This is naturally a subring of $F$, and we denote this ring as $S^{-1}R$.
\end{Definition}
Note that $R\subset S^{-1}R$, since we have assumed that $1\in S$.
Also note that if $S$ is generated by a finite set $\{s_1,\cdots, s_m\}$, then $S^{-1}R=R[s_1^{-1},\cdots, s_m^{-1}]$.
\begin{Lemma} \label{locallemma}
Let $\{p_1,p_2,\cdots,p_m\}$ and $\{q_1,q_2,\cdots ,q_m\}$ be two sets of independent variables with the following properties:
\begin{eqnarray}
p_j&\in&\mathbb{Z}\left[ q_1^{\pm}, q_2^{\pm},\cdots ,q_m^{\pm}\right], \label{p}\\
q_j&\in&\mathbb{Z}\left[ p_1^{\pm}, p_2^{\pm},\cdots ,p_m^{\pm}\right], \label{q}\\
q_j&&\mbox{is irreducible as an element of}\ \mathbb{Z}\left[ p_1^{\pm}, p_2^{\pm},\cdots ,p_m^{\pm}\right], \notag
\end{eqnarray}
for $j=1,2,\cdots, m$.
Let us take an irreducible Laurent polynomial
\[
f(p_1,\cdots,p_m)\in \mathbb{Z}\left[ p_1^{\pm}, p_2^{\pm},\cdots ,p_m^{\pm}\right],
\]
and another Laurent polynomial
\[
g(q_1,\cdots, q_m) \in \mathbb{Z}\left[ q_1^{\pm}, q_2^{\pm},\cdots ,q_m^{\pm}\right],
\]
which satisfies $f(p_1,\cdots,p_m)=g(q_1\cdots, q_m)$.
In these settings, the function $g$ is decomposed as
\[
g(q_1,\cdots, q_m)=p_1^{r_1}p_2^{r_2}\cdots p_m^{r_m}\cdot \tilde{g}(q_1,\cdots,q_m),
\]
where $r_1,r_2, \cdots, r_m\in\mathbb{Z}$ and $\tilde{g}(q_1,\cdots,q_m)$ is irreducible in $\mathbb{Z} \left[ q_1^{\pm}, q_2^{\pm},\cdots ,q_m^{\pm}\right]$.
\end{Lemma}
\textbf{Proof}\;\;
We use notations of multi-indices as $\boldsymbol{p}=(p_1,\cdots, p_m)$, $\boldsymbol{q}=(q_1,\cdots, q_m)$, and abbreviations  $\mathbb{Z}[p^{\pm}]=\mathbb{Z}[p_1^{\pm}, p_2^{\pm},\cdots , p_m^{\pm}]$, $\mathbb{Z}[q^{\pm}]=\mathbb{Z}[q_1^{\pm}, q_2^{\pm},\cdots ,q_m^{\pm}]$.
Let us suppose that $g$ can be factored as
\[
g(\boldsymbol{q})=g_1 (\boldsymbol{q}) g_2(\boldsymbol{q}),
\]
where $g_1,g_2\in \mathbb{Z}[q^{\pm}]$. Then we can prove that at least one of $g_1$ or $g_2$ should be a unit.
We can write
\[
g_1(\boldsymbol{q})=\frac{1}{\boldsymbol{q}^{\boldsymbol{s}}} P_1(\boldsymbol{q}),\;\;
g_2(\boldsymbol{q})=\frac{1}{\boldsymbol{q}^{\boldsymbol{t}}} P_2(\boldsymbol{q}),
\]
where $P_1,P_2$ are polynomials and the exponents of multi-indices are $\boldsymbol{s,t} \in\mathbb{Z}^m$.
If we transform the variables from $\boldsymbol{q}$ to $\boldsymbol{p}$, we have
\[
P_i(\boldsymbol{q})=Q_i(\boldsymbol{p})\in \mathbb{Z}[p^{\pm}]\;\;(i=1,2),
\]
which are no longer polynomials yet still Laurent polynomials,
since each $q_j$ is in $\mathbb{Z}[p^{\pm}]$ and $P_1,P_2$ are polynomials.
Therefore we have
\[
f(\boldsymbol{p}) = \frac{1}{\boldsymbol{q}^{\boldsymbol{s}+\boldsymbol{t}}} Q_1(\boldsymbol{p}) Q_2(\boldsymbol{p}).
\]
Since the function $f(\boldsymbol{p})$ is in $\mathbb{Z}[p^{\pm}]$ and each $q_j$ is irreducible in $\mathbb{Z}[p^{\pm}]$, the term
$\boldsymbol{q}^{\boldsymbol{s}+\boldsymbol{t}}$ must divide $Q_1(\boldsymbol{p}) Q_2(\boldsymbol{p})$.
Thus $Q_1$ and $Q_2$ have monomial factors of $\boldsymbol{q}$ as
\[
Q_1(\boldsymbol{p})=\boldsymbol{q}^{\boldsymbol{s}'} \cdot \tilde{Q}_1(\boldsymbol{p}),\; \; Q_2(\boldsymbol{p})=\boldsymbol{q}^{\boldsymbol{t}'} \cdot \tilde{Q}_2(\boldsymbol{p}),
\]
where $\boldsymbol{s}+\boldsymbol{t}=\boldsymbol{s}'+\boldsymbol{t}'$, $\tilde{Q}_i\in \mathbb{Z}[p^{\pm}]$.
Because of the irreducibility of $f(\boldsymbol{p})$ in $\mathbb{Z}[p^{\pm}]$, at least one of $\tilde{Q}_1$ or $\tilde{Q}_2$ must be a unit in $\mathbb{Z}[p^{\pm}]$ (of the form $\boldsymbol{p}^{\boldsymbol{r}}$, $\boldsymbol{r}\in\mathbb{Z}^m$).
We can assume that $\tilde{Q}_1$ is a unit, and we obtain
\[
g_1(\boldsymbol{q})=\boldsymbol{p}^{\boldsymbol{r}}\cdot \boldsymbol{q}^{\boldsymbol{s}'-\boldsymbol{s}},
\]
where $\boldsymbol{p}^{\boldsymbol{r}}=p_1^{r_1} p_2^{r_2} \cdots p_m^{r_m}$, $\boldsymbol{q}^{\boldsymbol{s}'-\boldsymbol{s}}=q_1^{s'_1-s_1} q_2^{s'_2-s_2} \cdots q_m^{s'_m-s_m}$.
Thus the lemma is proved. Note that $q_j^{s'_j-s_j}$ is a unit in $\mathbb{Z}[q^{\pm}]$.
\qed
Note that the conditions \eqref{p} and \eqref{q} in the lemma \ref{locallemma} are equivalent to the following fact:
two localized rings $\mathbb{Z}\left[ p_1^{\pm}, p_2^{\pm},\cdots ,p_m^{\pm}\right][q_1^{-1},q_2^{-1},\cdots,q_m^{-1}]$ and $\mathbb{Z}\left[ q_1^{\pm}, q_2^{\pm},\cdots ,q_m^{\pm}\right][p_1^{-1},p_2^{-1},\cdots,p_m^{-1}]$ are identical as subrings of $\mathbb{Q}(p_1,\cdots, p_m)$ if we substitute $q_j$'s for $p_j$'s by \eqref{q}.
Therefore the statement in lemma \ref{locallemma} is essentially a special version of the lemma \ref{loc} in the next section, which states that irreducibility is preserved under localization. However, we do not need the lemma \ref{loc} in this section, and we shall return to this point in the next section when we deal with the discrete KdV equation.
\begin{Theorem} \label{somosirred}
The term $y_n$ in Somos-4 \eqref{somos4} is irreducible as a Laurent polynomial in the ring 
$R:=\mathbb{Z}\left[a^{\pm},b^{\pm},c^{\pm},d^{\pm}\right]$.
Moreover, every pair of two terms $y_n,y_m (n\neq m)$ is co-prime in $R$.
\end{Theorem}
\textbf{Proof}\;\;
Let us define a sequence of integers $\{s_n\}_{n=1}^{\infty}$ as a Somos-4 sequence for initial values $y_1=y_2=y_3=y_4=1$: i.e.,
\[
s_n:=y_n|_{a=b=c=d=1},
\]
\[
\{s_n\}_{n=1}=\{1,1,1,1,2,3,7,23,59,314,1529,8209,83313,620297,\cdots\}.
\]
We prove, by induction, the irreducibility of $y_n$ for $n\ge 6$, since the case of $n=1,2,3,4,5$
is trivial.
\begin{itemize}
\item The case of $n=6,7,8,9$:
We define another ring of Laurent polynomials $R':=\mathbb{Z}[b^{\pm},c^{\pm},d^{\pm},y_5^{\pm}]$.
Let us suppose that $y_m (m<n)$ is irreducible in $R$ and let us prove the irreducibility of $y_n$.
Using the induction hypothesis that $y_{n-1}$ is irreducible in $R$, and by shifting the subscripts ($n-1\to n$, $a\to b$, $b\to c$, $c\to d$, $d\to y_5$), we have that $y_n$ is an irreducible element of $R'$.
We use lemma \ref{locallemma} in the case of $m=4$,
\[
(p_1,p_2,p_3,p_4)=(b,c,d,y_5),\ \ (q_1,q_2,q_3,q_4)=(a,b,c,d).
\]
We easily verify that $a,b,c,d\in R'$, $y_5\in R$, and $y_5$ is irreducible in $R$, since $y_5=(bd+c^2)/a$.
Since $y_n=:f(b,c,d,y_5)$ is irreducible in $R'$, the same function $y_n=:g(a,b,c,d)$ expressed in terms of the generators of $R$ (obtained by substituting $y_5=(bd+c^2)/a$), is decomposed as
\[
y_n=b^{r_1} c^{r_2} d^{r_3} y_5^{r_5} \cdot \tilde{g}(a,b,c,d),
\]
where $r_1,r_2,r_3,r_5\in\mathbb{Z}$ and $\tilde{g}(a,b,c,d)\in R$ is irreducible in $R$.
Since $b^{r_1}c^{r_2}d^{r_3}$ is a monomial (a unit in the ring $R$), $F:=b^{r_1}c^{r_2}d^{r_3}\cdot \tilde{g}$ is irreducible in $R$.
From $y_n\in R$ by proposition \ref{somosLaurent}, the exponent $r_5$ cannot be negative: $r_5\ge 0$. 
By substituting $a=b=c=d=1$, we obtain
\[
s_n=\left(F|_{a=b=c=d=1}\right)\cdot 2^{r_5}.
\]
Since $s_n$ is an odd integer for $n=6,7,8,9$, and $F|_{a=b=c=d=1}\in\mathbb{Z}$,
the exponent $r_5$ must be zero. Therefore $y_n (n=6,7,8,9)$ is irreducible in $R$.
\item The case of $n=10$:
Since $s_{10}=314$ is even, we have to take an approach different from the previous one.
Let us define a new Laurent ring $R'':=\mathbb{Z}[y_6^{\pm}, y_7^{\pm}, y_8^{\pm}, y_9^{\pm}]$.  
We use lemma \ref{locallemma} in the case of $m=4$,
\[
(p_1,p_2,p_3,p_4)=(y_6,y_7,y_8,y_9),\ \ (q_1,q_2,q_3,q_4)=(a,b,c,d).
\]
Let us take $f(y_6,y_7,y_8,y_9):=y_{10}=(y_7 y_9+y_8^2)/y_6$, which is irreducible in $R''$.
From proposition \ref{somosLaurent}, we have $y_6,y_7,y_8,y_9\in R$. Also we have $a,b,c,d\in R''$ ($a,b,c,d$ are Laurent polynomials of $y_j (j=6,7,8,9)$), since Somos-4 sequence \eqref{somos4} is reversible.
Note that $a,b,c,d$ are all irreducible in $R''$, which is obtained from the induction hypothesis and from the reversibility of Somos-4 sequence. Thus we have that all the conditions of lemma \ref{locallemma} are satisfied.
Therefore $y_{10}$ can be decomposed as an element of $R''$ as
\begin{equation}
y_{10}=y_6^{r_6}y_7^{r_7}y_8^{r_8}y_9^{r_9}\cdot G, \label{y10}
\end{equation}
where $G$ is irreducible in $R$ and $r_j$ is an integer for $j=6,7,8,9$.
By the Laurentness of $y_{10}$ from proposition \ref{somosLaurent}, we have $r_j\ge 0$ for $j=6,7,8,9$.
By substituting $a=b=c=d=1$ to \eqref{y10} we have
\[
314=3^{r_6} 7^{r_7} 23^{r_8} 59^{r_9} \cdot G|_{a=b=c=d=1} .
\]
Since $G|_{a=b=c=d=1}\in\mathbb{Z}$, we have $r_j=0$ for $j=6,7,8,9$.
Thus $y_{10}$ is irreducible in $R$.
\item The case of $n=11,12,13$ is proved by the same argument as in the case of $n=6,7,8,9$.
\item The case of $n\ge 14$:
We define another ring $R''':=\mathbb{Z}[y_{10}^{\pm},y_{11}^{\pm},y_{12}^{\pm},y_{13}^{\pm}]$.
By the induction hypothesis, $y_n$ is irreducible in all the rings $R'$, $R''$ and $R'''$.
Using lemma \ref{locallemma}, like we have done in the case of $n=10$, we have
\begin{eqnarray*}
y_n&=&F\cdot y_5^{r_5}\\
&=&G\cdot y_6^{r_6}y_7^{r_7}y_8^{r_8}y_9^{r_9}\\
&=&H\cdot y_{10}^{r_{10}}y_{11}^{r_{11}}y_{12}^{r_{12}}y_{13}^{r_{13}},
\end{eqnarray*}
where $r_j$ is a non-negative integer and $F,G,H$ are irreducible in $R$.
From numerical computation we can see that $y_5,\cdots, y_{13}$ are all distinct from each other. Therefore each pair of two terms $y_n,y_m$ with $5\le n,m\le 13$ is co-prime.
From the second equality and the irreducibility of $G$ in $R$,
we have that $r_5=0$ or $r_5=1$.
If $r_5=1$, then $G=y_5\cdot u$ where $u$ is a unit in $R$.
Then the third equality and the irreducibility of $H$ in $R$ indicate that $r_6=\cdots=r_{13}=0$ and that $F=u$. Thus $y_n/y_5$ has to be a unit in $R$. We write $y_n=y_5\cdot v$, where $v$ is a unit in $R$. Substituting $a=b=c=d=1$ yields $s_n=s_5=2$. Since $s_n$ is strictly increasing, this is a contradiction.
Therefore we conclude that $r_5=0$.
Therefore $y_n=F$ is irreducible in $R$.
\end{itemize}
Hence, the irreducibility of $y_n$ over $R$ has been proved.
By the irreducibility of $y_n$ as a Laurent polynomial, the numerator of $y_n$ (which we denote by $z_n$) is irreducible as a polynomial. We easily see that, since each $z_n$ has a distinct degree for distinct $n\ge 5$,
$z_n$ and $z_m$ are co-prime as polynomials for $n\neq m$.
We have proved that two terms of the Somos-4 sequence $y_n,y_m$ are co-prime if $n\neq m$. \qed
The irreducibility of $y_n (n\ge 6)$ over $k[a^{\pm},b^{\pm},c^{\pm},d^{\pm}]$, where $k$ is any field of characteristic zero can be proved using theorem \ref{somosirred} and several facts from algebra.
This is explained in lemma \ref{charzero}.
\begin{Definition}
The Laurent polynomial with integer coefficients $f=\sum a_J \boldsymbol{z}^J\in\mathbb{Z}[z_1^{\pm},\cdots, z_n^{\pm}]$  ($\boldsymbol{z}^J$ is a multi-index notation)
is said to be primitive if gcd$_J(a_J)=1$.
\end{Definition}
\begin{Lemma}[Gauss lemma] \label{gausslem}
Let us take Laurent polynomials with integer coefficients as
\[
f,g\in \mathbb{Z}[z_1^{\pm},\cdots, z_n^{\pm}].
\]
Let us suppose that $f$ is primitive and that there exists $h\in \mathbb{Q}[z_1^{\pm},\cdots, z_n^{\pm}]$ with rational coefficients
such that $g=fh$.
Then we have $h \in\mathbb{Z}[z_1^{\pm},\cdots, z_n^{\pm}]$.
\end{Lemma}
\begin{Lemma} \label{charzero}
Let us take a field $k$ of characteristic zero.
The terms of the Somos-4 sequence \eqref{somos4} $y_n$ are irreducible elements in the ring $\tilde{R}:=k[a^{\pm},b^{\pm},c^{\pm},d^{\pm}]$,
where $y_0=a,y_1=b,y_2=c,y_3=d$.  
\end{Lemma}
\textbf{Proof} \;\;
We show this lemma by induction.
Let us take another ring $\tilde{R}':=k[b^{\pm},c^{\pm},d^{\pm}, y_5^{\pm}]$.
By the induction hypothesis, the term $y_n$ is irreducible in the ring $\tilde{R}'$.
From the relation of localized rings $\tilde{R}'\subset \tilde{R}'[a^{-1}]=\tilde{R}[y_5^{-1}]$,
we have that $y_n$ is also irreducible in $\tilde{R}[y_5^{-1}]$.
Thus we can express $y_n$ as $y_n=F\cdot y_5^r$, where $F$ is irreducible in $\tilde{R}$ and $r\ge 0$.
Since $y_n,y_5\in R=\mathbb{Z}[a^{\pm},b^{\pm},c^{\pm},d^{\pm}]$ holds
from proposition \ref{somosLaurent}, it follows that $F\in\mathbb{Q}[a^{\pm},b^{\pm},c^{\pm},d^{\pm}]$.
Since $y_5=\frac{bd+c^2}{a}$ is primitive, we conclude from lemma \ref{gausslem} that $F\in R$.
By the irreducibility of $y_n$ over $R$ already obtained in theorem \ref{somosirred}, we have $r=0$ and $y_n=F$.
\qed
Finally we can prove the assertions concerning the factors of $x_n$ of the QRT mapping \eqref{QRT2} using the properties of the Somos-4 sequence \eqref{somos4}.
\begin{Theorem} \label{QRTprop}
For the QRT mapping \eqref{QRT2}, terms $x_n\in\mathbb{C}(t,u)$ have the following property:
For every integer $l\ge 2$, there exists a unique polynomial $F_l\in\mathbb{Z}[t,u]$ such that
\[
\mbox{ord}_{F_l}x_l=1,\; \mbox{ord}_{F_l}x_{l+1}=-2,\; \mbox{ord}_{F_l}x_{l+2}=1,
\]
and that ord$_{F_l} x_m=0$ for every $m\neq l,l+1,l+2$, which is not a monomial.
\end{Theorem}
\textbf{Proof}\;\;
Let us study the correspondence between the initial values $a,b,c,d$ of the Somos-4 sequence \eqref{somos4}, and those of the QRT mapping \eqref{QRT2}.
Since $x_0=u=\dfrac{ac}{b^2}$ and $x_1=t=\dfrac{bd}{c^2}$, we have
\[
c=\frac{b^2}{a}u,\; d=\frac{b^3}{a^2}t u^2.
\]
Therefore the transformation from $\{a,b,c,d\}$ to $\{a,b,t,u \}$ is expressed by means of monomial relations.
Thus, using theorem \ref{somosirred}, we have that
\[
y_n\in\mathbb{Z}[a^{\pm},b^{\pm},t^{\pm},u^{\pm}],
\]
and that $y_n$ is irreducible as a Laurent polynomial, and furthermore that co-primeness holds in $\mathbb{Z}[a^{\pm},b^{\pm},t^{\pm},u^{\pm}]$.
By the fact that $y_n$ has the form
\[
y_n=\frac{b^{n-1}}{a^{n-2}}z_n,\;\; z_n\in\mathbb{Z}[u^{\pm}, t^{\pm}],
\]
and by the relation \eqref{somosrel}, we confirm that $x_n\in\mathbb{Q}(u,t)$, i.e., the $x_n$ do not depend on $a$ or $b$.
Since we need only three terms $y_{n+1},y_{n+2},y_{n+3}$
to define $x_n$ from \eqref{somosrel}, we obtain the property that only three consecutive terms have a common factor other than monomials of $u$ and $t$, either in their denominators or in their numerators.
Valuations of $x_n$ with respect to the common factor are readily obtained.
\qed
The singularity of this equation \eqref{QRT2} is confined if we evolve the equation at least for three steps, because the number of terms that have a particular polynomial $F_l$ as a common factor, is limited to three.

Next let us compare our previous results with that of the following non-integrable equation:
\begin{equation}
x_{n+1}=\frac{x_n +1}{x_{n-1}x_n^3}, \label{nonQRT3}
\end{equation}
with the initial condition $x_1=u,x_2=t$.
The evolution of the equation \eqref{nonQRT3} is
\begin{eqnarray*}
x_2&=&\frac{1+t}{t^3 u},x_3=\frac{t^5 u^2(1+t+t^3 u)}{(1+t)^3},x_4=\frac{(1+t)^5((1+t)^3+t^5 u^2+t^6 u^2+t^8 u^3)}{t^{12} u^5(1+t+t^3 u)^3},\\
x_5&=&\frac{t^{19} u^8(1+t+t^3 u)^5(1+8t+\cdots+3t^{19} u^7+t^{21} u^8)}{(1+t)^{12}((1+t)^3+t^5 u^2+t^6 u^2+t^8 u^3)^3},\cdots.
\end{eqnarray*}
In this case, the valuation of $x_m$ with respect to $(1+t)$ is non-zero for $x_m(m\ge 2)$, i.e.,
\[
\mbox{ord}_{1+t}x_2=1,\ \mbox{ord}_{1+t}x_3=-3,\ \mbox{ord}_{1+t}x_4=5,\ \mbox{ord}_{1+t}x_5=-12,\cdots.
\]
In fact, we can prove that $|$ord$_{(1+t)}x_m|\to \infty (m\to \infty)$.
Let us define $d_m:=| $ord$_{(1+t)}x_m |$.
We can prove inductively that $d_m$ satisfies
\[
d_{m+2}=\left\{
\begin{array}{cl}
2d_{m+1}-d_m & (m=2,4,6,\cdots)\\
3d_{m+1}-d_m & (m=3,5,7,\cdots)
\end{array},
\right.
\]
with $d_2=1,d_3=3$.
From this recurrence relation we have $\lim_{m\to\infty} d_m=\infty$.
We obtain exactly the same recurrence for the absolute values of the orders of other common factors (e.g. $|$ord$_{1+t+t^3 u} x_m |$).
These facts indicate that $x_m$ and $x_n$ are never co-prime (i.e., they have common factors other than monomials) for $m,n\ge 2,\, m\neq n$.
Therefore the singularities are not confined in the equation \eqref{nonQRT3}.
In this section, we have observed that, at least for equations \eqref{QRT2} and \eqref{nonQRT3},  the integrability detection using the co-primeness of the several consecutive terms gives the same result as the singularity confinement test.

\section{Co-primeness of discrete KdV equation} \label{section3}
In this section we extend our previous discussion to partial difference equations, in particular, to the discrete KdV equation.
Here we take the following initial data for equations \eqref{nonlin} and \eqref{bilin}:
\begin{eqnarray*}
I_w&=&\{w_m^0,\, w_1^n\ |\ m\ge 1,n\ge 0\},\\
I_a&=&\{a_m^0=1,\, a_0^1=1,\, a_m^1=x_m,\, a_0^n=y_n\ |\ m\ge 1, n\ge 2\}.
\end{eqnarray*}
The set $I_w$ is the most general initial condition for nonlinear dKdV \eqref{nonlin}. Elements of $I_a$ will be taken to be compatible with $I_w$. Then the correspondence between $I_a$ and $I_w$ is:
\begin{equation}
x_l=\left(\prod_{i=1}^l w_i^0\right)^{-1}\ (l\ge 1),\ y_l=\left(\prod_{j=1}^{l-1} \alpha_j\right)^{-1}\ (l\ge 2), \label{xy1}
\end{equation}
where
\begin{equation}
\alpha_l=\prod_{j=1}^l \beta_j\ (l\ge 1),\ \beta_j=\frac{1+\delta}{w_1^j}-\delta w_1^{j-1}\ (j\ge 0). \label{ab1}
\end{equation}
We briefly explain how to obtain \eqref{xy1}.
First we prove the expression for $x_l$. We have from \eqref{relation},
\[
w_1^0=\frac{a_0^1 a_1^0}{a_1^1 a_0^0}=\frac{1}{x_1},\; w_2^0=\frac{a_1^1 a_2^0}{a_2^1 a_1^0}=\frac{x_1}{x_2},\cdots.
\]
Thus $x_l=\left(\prod_{i=1}^l w_i^0\right)^{-1}$ is proved for $l\ge 1$.
Next we prove the expression for $y_l$. From the equation \eqref{bilin} and the relation \eqref{relation}, we have
\[
(1+\delta) y_{k-1} a_1^{k+1} = y_k a_1^k +\delta y_{k+1} a_1^{k-1},
\]
and
\[
w_1^k=\frac{y_{k+1} a_1^k}{y_k a_1^{k+1}}.
\]
From these two equations we have
\[
\frac{1+\delta}{w_1^k}-\delta w_1^{k-1}=\frac{y_k^2}{y_{k-1} y_{k+1}}.
\]
Thus if we take $\alpha_j, \beta_j$ as in \eqref{ab1} we obtain
\[
\alpha_k=\prod_{l=1}^k \beta_l=\frac{y_k}{y_{k+1}},
\]
where we have formally taken $y_0:=a_0^0=1, y_1:=a_0^1=1$.
Therefore we have proved that
\[
y_n=\frac{y_1}{\prod_{k=1}^{n-1} \alpha_k}=\frac{1}{\prod_{k=1}^{n-1} \alpha_k}.
\]
Let us introduce the parameter
\[
\gamma_j:=w_1^{j-1}w_1^j-\frac{1+\delta}{\delta}=-\frac{1}{\delta}w_1^j \beta_j,
\]
for convenience.
\begin{Theorem} \label{thm1}
Let $K$ be a field, and let us suppose that $\delta\neq 0,-1$.
Then the term $a_m^n$ of the bilinear dKdV equation \eqref{bilin} is a Laurent polynomial of the initial variables, i.e.,
\[
a_m^n\in K[(x_1)^{\pm}, (x_2)^{\pm},\cdots, (y_2)^{\pm}, (y_3)^{\pm},\cdots].
\]
We also have that $a_m^n$ is an irreducible Laurent polynomial and that $a_m^n$ and $a_{m'}^{n'}$
are co-prime if $(m,n)\neq (m',n')$.
\end{Theorem}
\textbf{Proof} The proof can be found in \cite{dKdVSC}.\qed
Note that we do not need any restriction on the characteristic of the field $K$.
Let us define the substitution map:
\[
\Phi: K[(x_1)^{\pm},(x_2)^{\pm},\cdots,(y_2)^{\pm},(y_3)^{\pm}\cdots] \to K(w_1^0,w_2^0\cdots, w_1^1,w_1^2,\cdots).
\]
The image $\Phi(a_m^n)$ is obtained by substituting the variables $x_1,x_2,\cdots,y_2,y_3\cdots$
 with $w_1^0,w_2^0\cdots, w_1^1,w_1^2,\cdots$ by \eqref{xy1} and \eqref{ab1}.
We denote the image of $\Phi$ as $A$.
We define another ring $B$:
\[
B:=K[(w_1^0)^{\pm}, (w_2^0)^{\pm},\cdots, (w_1^1)^{\pm}, (w_1^2)^{\pm}, \cdots, (\gamma_1)^{\pm}, (\gamma_2)^{\pm},\cdots].
\]
We readily obtain $A\subset B$, since each $\alpha_k$ can be expressed using $\gamma_k$ and $w_1^k$.
In fact we have
\[
A(=\mbox{Im}\, \Phi)=K[(w_1^0)^{\pm}, (w_2^0)^{\pm}, \cdots, (\beta_1)^{\pm}, (\beta_2)^{\pm}, \cdots],
\]
and, hence, the mapping $\Phi$ is injective.
For a localization, as defined in definition \ref{localdef}, we have the following lemma:
\begin{Lemma} \label{loc2}
The extension of rings $A\subset B$ is a localization.
\end{Lemma}
\textbf{Proof}\;\;
First we have $(w_1^1)^{-1}\in A$, since $A\ni \beta_1+\delta w_1^0=\frac{1+\delta}{w_1^1}$.
We also have $\beta_2+\delta w_1^1=\frac{1+\delta}{w_1^2}$. Therefore $\frac{1}{w_1^1 w_1^2}=\frac{1}{1+\delta}\left(\frac{\beta_2}{w_1^1}+\delta \right)\in A$. In the same manner, we obtain
\[
\left(\prod_{j=1}^l w_1^j \right)^{-1}\in A,
\]
for $l=1,2,\cdots$.
The following set $S$ is multiplicative:
\[
S:=\left\{ c_0\left(\frac{1}{w_1^1}\right)^{c_1}\left(\frac{1}{w_1^2}\right)^{c_2}\cdots \left(\frac{1}{w_1^n}\right)^{c_n}\ \Big|\ n\in\mathbb{N},c_0\in K\setminus \{0\}, c_j\ge 0 \right\}.
\]
Therefore we can take the localization $S^{-1} A$, whose elements are written as $a/s$ with $s\in S$, $a\in A$.
We prove that
\[
B=S^{-1}A.
\]
Since $S^{-1} A\subset B$ is trivial, we only have to prove $B\subset S^{-1} A$. In fact, we have only to prove that $\gamma_j\in S^{-1} A$ for $j\ge 1$. From the construction of $S^{-1} A$, we have $w_1^j=1/(1/w_1^j)\in S^{-1} A$. Thus $\gamma_j=-\frac{1}{\delta} w_1^j \beta_j \in S^{-1} A$.
\qed
\begin{Lemma} \label{loc}
Let us take a unique factorization domain (UFD) $R$.  The localization $S^{-1} R$ of $R$ with respect to a multiplicative set $S$ conserves the irreducibility of the elements: i.e., if we take a non-zero irreducible element $p\in R$, $p$ is also irreducible in $S^{-1} R$. 
\end{Lemma}
\textbf{Proof}\;\; We can suppose that $p$ is not a unit in $R$, since a unit in $R$ is obviously a unit in $S^{-1} R$.
We prove by contradiction. Let us suppose that $p$ can be factored as $\displaystyle p=\frac{x}{s}\cdot \frac{y}{t}$, where $x,y\in R$ and $s,t\in S$. Then $p$ divides $xy$ in the ring $R$. Since $p$ is irreducible in $R$ and is not a unit of $R$, we have that $p$ divides either $x$ or $y$. We can suppose that $p$ divides $x$. Then there exists $z\in R$ such that $pz=x$. Then we have $pst=pzy$ and from the fact that $R$ is a domain, we obtain $yz=st$. Therefore $\displaystyle \frac{y}{t}\cdot \frac{z}{s}=1$ in the localized ring $S^{-1} R$, which indicates that $y/t$ is a unit in $S^{-1} R$. Thus $p$ is irreducible in $S^{-1} R$. \qed
%
\begin{Proposition} \label{prop1}
Let us suppose that $\delta\neq 0,-1$.
Each term in the bilinear dKdV equation \eqref{bilin} is an irreducible Laurent polynomial as an element of the localized ring $B$. Two terms $a_m^n$ and $a_{m'}^{n'}$ with $(m,n)\neq (m',n')$ are co-prime as elements of $B$, i.e., they do not have common factors other than monomials in $B$.
\end{Proposition}
\textbf{Proof}\;\;
Theorem \ref{thm1} indicates that the term $a_m^n$ is an irreducible Laurent polynomial as an element of the ring $K[(x_1)^{\pm},(x_2)^{\pm},\cdots,(y_2)^{\pm},(y_3)^{\pm}\cdots]$ of Laurent polynomials, and that two distinct terms are co-prime in this ring. Parameters $\beta_j$ are written as monomials of $y_1,\cdots ,y_j$ as $\displaystyle \beta_k=\frac{y_k^2}{y_{k-1} y_{k+1}}\ (k\ge 2)$. Therefore we have that $a_m^n$ is an irreducible Laurent polynomial as an element of the ring $A=$Im$\,\Phi$.
From lemmas \ref{loc2} and \ref{loc}, and from the fact that ring $A$ is a UFD, we conclude that $a_m^n$ is an irreducible Laurent polynomial also in the ring $B$, and that the co-primeness of distinct terms is true in the ring $B$. \qed
To apply this result to the nonlinear dKdV equation, we first define a notion analogous to co-primeness in the field of rational functions.
\begin{Definition}
Two rational functions $f$ and $g$ are said to be `co-prime' with respect to the ring of Laurent polynomials
$K[z_1^{\pm}, z_2^{\pm}, z_3^{\pm}, \cdots]$ if the following condition is satisfied:
Let us write $f=\displaystyle \frac{f_1}{f_2}$ and $g=\displaystyle \frac{g_1}{g_2}$, where
$f_1,f_2,g_1,g_2$ are polynomials, and the denominator and numerator of $f$ and $g$ do not share a factor.
Then every pair of polynomials among the four polynomials $f_1,f_2,g_1,g_2$ does not have any common factors other than monomials in terms of $z_1,z_2,z_3,\cdots$.
\end{Definition}
From proposition \ref{prop1}, we obtain the following proposition.
\begin{Proposition} \label{thm2}
Let us suppose that $\delta\neq 0,-1$.
Two terms $w_m^n$ and $w_{m'}^{n'}$ of the nonlinear dKdV equation \eqref{nonlin}
are `co-prime' with respect to the ring $B$ of Laurent polynomials on condition that
\[
|n-n'|\ge 2\ \mbox{or}\ |m-m'|\ge 2.
\]
\end{Proposition}
\textbf{Proof}\;\;
By examining the relation \eqref{relation}, we know that $w_m^n$ and $w_{m'}^{n'}$ use the same term $a_k^l$ of the bilinear dKdV equation, only when they are adjacent to each other. \qed
Next we investigate the same equation but with variables $m$ and $n$ interchanged.
Note that the equation
\[
\frac{1}{w_{m+1}^{n+1}}-\frac{1}{w_m^n}+\frac{\tilde{\delta}}{1+\tilde{\delta}}(w_{m+1}^n-w_m^{n+1})=0
\]
is equivalent to the original nonlinear dKdV equation \eqref{nonlin}, if we take
\[
\tilde{\delta}=\frac{-\delta}{1+2\delta}.
\]
We define a new transformation
\begin{equation}
w_m^n=\frac{\tilde{a}_m^{n-1} \tilde{a}_{m-1}^n}{\tilde{a}_m^n \tilde{a}_{m-1}^{n-1}}. \label{newrelation}
\end{equation}
The new variable $\tilde{a}_m^n$ satisfies the same form as that of bilinear dKdV \eqref{bilin}, but the variables $m$ and $n$ are interchanged:
\begin{equation}
(1+\tilde{\delta})\tilde{a}_{m-1}^{n-1} \tilde{a}_{m+1}^{n} =\tilde{\delta} \tilde{a}_{m+1}^{n-1} \tilde{a}_{m-1}^n +\tilde{a}_{m}^{n} \tilde{a}_m^{n-1}. \label{newbilin}
\end{equation}
This equation evolves in a same manner as \eqref{bilin} when we consider it on the $n$-$m$ plane, not $m$-$n$ plane.
We take the initial variables of the equation \eqref{newbilin} as
\[
I_{\tilde{a}}:=\{\tilde{a}_0^n=1, \tilde{a}_1^n=\tilde{x}_n, \tilde{a}_m^0=\tilde{y}_m, \tilde{a}_1^0=1\ |\ m\ge 2, n\ge 1\}.
\]
With calculations similar to those derived \eqref{xy1} and \eqref{ab1}, we obtain the following relations between the initial values of $w_m^n$ and $\tilde{a}_m^n$:
\begin{equation}
\tilde{x}_l=\left(\prod_{i=1}^l w_1^i\right)^{-1}\ (l\ge 1),\ \tilde{y}_l=\left(\prod_{j=1}^{l-1} \tilde{\alpha}_j\right)^{-1}\ (l\ge 2),
\end{equation}
where
\begin{equation}
\tilde{\alpha}_l=\prod_{j=1}^l \tilde{\beta}_j\ (l\ge 1),\ \tilde{\beta}_j=\frac{1+\tilde{\delta}}{w_{j+1}^0}-\tilde{\delta} w_j^{0}\ (j\ge 0).
\end{equation}
We define parameters
\[
\tilde{\gamma}_j:=w_j^0 w_{j+1}^0-\frac{1+\tilde{\delta}}{\tilde{\delta}}\ (j=0,1,\cdots),
\]
and define the ring $\tilde{B}$ as
\[
\tilde{B}:=K[(w_1^0)^{\pm},(w_2^0)^{\pm},\cdots, (w_1^1)^{\pm}, (w_1^2)^{\pm}, \cdots, (\tilde{\gamma}_0)^{\pm},(\tilde{\gamma}_1)^{\pm},\cdots].
\]
We easily obtain the following statement, which is similar to proposition \ref{prop1}. 
\begin{Proposition}\label{prop3}
Let us suppose that $\tilde{\delta}\neq 0,-1$. The term
$\tilde{a}_m^n$ of the equation \eqref{newbilin} is an irreducible Laurent polynomial as an element of $\tilde{B}$, and the distinct terms are co-prime in $\tilde{B}$.
\end{Proposition}
Therefore we obtain the following proposition, which corresponds to proposition \ref{thm2}.
\begin{Proposition}\label{thm3}
Suppose that $\tilde{\delta}\neq 0,-1$.
Two terms $w_m^n$ and $w_{m'}^{n'}$ of the nonlinear dKdV equation \eqref{nonlin} are `co-prime' with respect to the ring $\tilde{B}$ of Laurent polynomials
 on condition that
\[
|n-n'|\ge 2\ \mbox{or}\ |m-m'|\ge 2.
\]
\end{Proposition}
From propositions \ref{thm2} and \ref{thm3} we conclude that
the term $w_m^n$ in nonlinear dKdV equation \eqref{nonlin} does not have common factors other than
monomials of $w_1^0,w_2^0,\cdots$ and $w_1^1,w_1^2,\cdots$.
Because the two sets of irreducible polynomials $\{\gamma_1,\gamma_2,\cdots\}$ and $\{\tilde{\gamma}_0,\tilde{\gamma}_1,\cdots\}$ do not have common elements, neither $\gamma_j$ nor $\tilde{\gamma}_j$ can be a factor of $w_m^n$. Thus we obtain our main theorem:
\begin{Theorem} \label{thm4}
Let us suppose that $\delta\neq 0,-1,-1/2$.
Two terms $w_m^n$ and $w_{m'}^{n'}$ of the nonlinear dKdV equation \eqref{nonlin} are `co-prime'
with respect to the ring of Laurent polynomials 
\[
K[(w_1^0)^{\pm},(w_2^0)^{\pm},\cdots, (w_1^1)^{\pm}, (w_1^2)^{\pm},\cdots]
\]
generated by the initial variables, on condition that
\[
|n-n'|\ge 2\ \mbox{or}\ |m-m'|\ge 2.
\]
\end{Theorem}
The above theorem suggests that two terms of the equation \eqref{nonlin} do not have common factors in their denominators or numerators, if they are separated by more than one cell in at least one of the coordinates. This means that the poles and zeros of the nonlinear dKdV equation do not propagate beyond two steps away in each direction. This fact can be interpreted as the confinement of singularities of the equation, and therefore gives the precise description of the singularity confinement criterion for the equation \eqref{nonlin}.

This theorem is an extension of the result in our previous paper \cite{dKdVSC} to general boundary and initial conditions.
We also have the additional advantage that the characteristic of the field $K$ is not restricted. Therefore the theorems are
valid for equations over a finite field $K$ as long as $\delta$ is transcendental over $K$ (in this case, $\delta$ can be seen as another independent variable), or $\delta\neq -1,0,-1/2$ is satisfied.

%
\section{Concluding remarks and discussions}
In this paper we investigated the confinement of singularities of an integrable mapping related to the Somos-4 sequence and the discrete KdV equation in terms of the irreducibility and co-primeness of their terms.
First, in section \ref{section2}, we proved the co-primeness of the terms of the Somos-4 sequence \eqref{somos4} (theorem \ref{somosirred}). Using this theorem we proved that only three adjacent terms of the QRT mapping \eqref{QRT2} have common factors other than monomials of the initial variables (theorem \ref{QRTprop}). This `co-primeness' property can be considered to be one of the powerful mathematical representations of confinement of singularities.
In contrast to this result, the non-integrable mapping \eqref{nonQRT3} does not have such `co-primeness' property: i.e., infinitely many of the terms of \eqref{nonQRT3} have non-monomial common factors.
In section \ref{section3}, we have proved in proposition \ref{prop1} that the terms $a_m^n$ of the bilinear dKdV equation \eqref{bilin} are irreducible Laurent polynomials in terms of the initial variables, and that distinct terms are co-prime as Laurent polynomials.
From proposition \ref{prop1} we have proved in proposition \ref{thm2} that the terms $w_m^n$ and $w_{m'}^{n'}$ of the nonlinear dKdV equation \eqref{bilin}, if separated by more than one cell, do not have common factors other than monomials (of the initial data and a class of auxiliary variables $\{\gamma_i\}$.)
We then studied another expression of the bilinear dKdV equation \eqref{newbilin}. The equation \eqref{newbilin} evolves in a manner similar to that of \eqref{bilin}, yet the coordinates $n$ and $m$ are interchanged.
The equation \eqref{newbilin} gives the same nonlinear dKdV equation \eqref{nonlin} through the relation \eqref{newrelation}.
In proposition \ref{prop3} we proved the Laurentness, the irreducibility and co-primeness also for the equation \eqref{newbilin}. From this we have proposition \ref{thm3}, which is an analogue of the proposition \ref{thm2}.
By combining the two propositions \ref{thm2} and \ref{thm3} we obtained our main theorem \ref{thm4} that the terms of the nonlinear dKdV equation \eqref{nonlin} do not have common factors other than monomials made solely of initial values
$\{w_m^0,w_1^n|m\ge 0, n\ge 1\}$.

Our assertion is that the `co-primeness' is an alternative mathematical statement of singularity confinement for nonlinear partial difference equations, and that it gives an integrability criterion for such difference equations.
It is reasonable to assume that a similar discussion should be possible for various integrable and non-integrable equations. In fact, we are currently investigating the co-primeness of the discrete Toda equation for several boundary conditions, and we hope to report the results in the future. Applications of this approach to higher dimensional systems such as the Hirota-Miwa equation will be also studied in future works. 
It will be an interesting task to investigate the relation between integrability and co-primeness for chaotic systems such as the one treated in \cite{HV} and linearizable systems, for which the singularity confinement approach is not equivalent to the integrability of the equations.
We expect to obtain a unified integrability criterion for many of the discrete integrable equations, by investigating the relation of our method to other integrability criteria: algebraic entropy \cite{BV}, singularity confinement for ultra-discrete equations \cite{JL,Ormerod}, and the almost good reduction criterion \cite{KMTT}.

\section*{Acknowledgments}
The authors wish to thank Prof. R. Willox and Prof. P. H. van der Kamp for useful comments.
This work is partially supported by Grants-in-Aid for Scientific Research of JSPS ($24\cdot 1379$, $25\cdot 3088$, $26\cdot 242$), and the Program for Leading Graduate Schools, MEXT, Japan.

\end{document}